\documentclass[conference]{IEEEtran}
\IEEEoverridecommandlockouts

\usepackage{cite}
\usepackage{amsmath,amssymb,amsfonts}
\usepackage{algorithmic}
\usepackage{graphicx}
\usepackage{textcomp}
\usepackage{xcolor}
\usepackage{comment}
\usepackage{url}
\usepackage{booktabs}
\usepackage{algorithmic}
\usepackage{algorithm}
\usepackage{multirow}
    
\usepackage[pscoord]{eso-pic} 

\begin{document}

\title{A New Approach for Evaluating the Performance of Distributed Latency-Sensitive Services}

\author{\IEEEauthorblockN{Theodoros Theodoropoulos, John Violos, Antonios Makris, Konstantinos Tserpes}
\IEEEauthorblockA{\textit{Department of Informatics and Telematics, Harokopio University of Athens, Greece}
}
\IEEEauthorblockA{\textit{School of Electrical and Computer Engineering, National Technical University of Athens, Greece}
}
}


\maketitle

\begin{abstract}
Conventional latency metrics are formulated based on a broad definition of traditional monolithic services, and hence lack the capacity to address the complexities inherent in modern services and distributed computing paradigms. Consequently, their effectiveness in identifying areas for improvement is restricted, falling short of providing a comprehensive evaluation of service performance within the context of contemporary services and computing paradigms. More specifically, these metrics do not offer insights into two critical aspects of service performance: the frequency of latency surpassing specified Service Level Agreement (SLA) thresholds and the time required for latency to return to an acceptable level once the threshold is exceeded. This limitation is quite significant in the frame of contemporary latency-sensitive services, and especially immersive services that require deterministic low latency that behaves in a consistent manner. Towards addressing this limitation, the authors of this work propose 5 novel latency metrics that when leveraged alongside the conventional latency metrics manage to provide advanced insights that can be potentially used to improve service performance. The validity and usefulness of the proposed metrics in the frame of providing advanced insights into service performance is evaluated using a large-scale experiment.

\end{abstract}

\section{Introduction}
\label{intro}
Beyond 5G (B5G) services, such as eXtended Reality (XR) and Multiplayer Mobile Gaming (MMG), are frequently intertwined with a multitude of demanding Quality of Service (QoS) requirements. Both application classes rely on delivering an immersive experience for a diverse set of end-users. Achieving acceptable levels of immersion necessitates extremely low latency and high bandwidth \cite{10.1145/2938559.2938583}. Furthermore, the unavoidable occurrence of system failures can have serious consequences for the realization of immersive experiences, as they frequently lead to disruptions in service delivery, thereby jeopardizing the intended level of immersion. Consequently, these applications should exhibit characteristics that enable them to withstand and manage faults, ensuring a level of fault tolerance \cite{jalote1994fault}.

Another crucial requirement for XR and MMG services is the imperative for end-user equipment to be as lightweight and cost-effective as possible \cite{theodoropoulos2022cloud}. Although cloud computing can transfer the computational burden to remote resources, enabling end-user devices to be mobile and economical, it falls short in fully supporting immersive applications that demand low latency and high bandwidth because end-user devices are typically distant from the cloud servers. This disparity results in processing and network overheads, leading to suboptimal performance and increased latency. Edge computing aims to diminish the amount of data requiring transmission to remote clouds and facilitates data processing in close proximity to the data sources. Consequently, edge computing can offer faster response times, elevated transfer rates, and improved scalability, and availability. As a result, deploying this type of services in a distributed manner across the cloud-edge infrastructure would assist in meeting the aforementioned QoS requirements \cite{taleb2022towards}.

Auto-scaling is a critical aspect of both edge and cloud computing environments, ensuring efficient resource utilization and performance consistency amidst changing workloads. This dynamic provisioning process adjusts computational resources based on current demands. In cloud computing, it responds to fluctuating service demands by provisioning additional virtual machines or scaling down surplus resources to optimize performance and reduce costs \cite{singh2019research}. Similarly, in edge computing, auto-scaling is crucial, but it prioritizes distributing computational resources across edge nodes to minimize latency and enhance efficiency \cite{violos2022intelligent}.

Service Level Agreements (SLAs) \cite{luu_admission_2022} are essential for upholding agreed-upon service standards in the frame of both edge and cloud computing. These agreements may cover an extensive array of QoS metrics \cite{syu_qos_2021} that the provider commits to ensuring. These QoS metrics encompass aspects such as latency, throughput, fault tolerance, and energy consumption, among others that are explicitly stated in the form of specified thresholds. 

Over the years, there has been a plethora of latency metrics \cite{aslanpour2020performance} that were introduced in the frame of evaluating service performance. By regularly analyzing and interpreting these metrics, service providers can identify areas for improvement, and optimize performance, accordingly. However, since these metrics were conceptualized on the basis of a rather generic and broad definition of conventional monolithic services, they fail to cater to the intricacies of contemporary services and distributed computing paradigms. As a result, the range of areas for improvement that they can establish is limited and not capable of fully evaluating service performance in the frame of contemporary services and distributed computing paradigms. 

To address this limitation, the authors of this work propose $5$ novel latency metrics that are based on fault tolerance metrics and that when leveraged alongside the conventional latency metrics manage to provide advanced insights that can be potentially used to improve service performance. Towards achieving this goal, this paper is structured in the following manner. Section \ref{lat} describes the existing conventional latency metrics and analyzes their limitations. Section \ref{prop} explores various fault tolerance metrics, and showcases the proposed latency metrics that are based on the former. Section \ref{exp} describes the experimental procedures employed to assess the effectiveness of the suggested solution. Section \ref{conclusion} encapsulates the merits and discoveries of this study.

\section{Latency Metrics}
\label{lat}
Latency \cite{srivastava2016survey} and execution time \cite{bielecki2022estimation} are intimately connected in the realm of computing, with latency representing the time required for the execution of a program or task to commence, and execution time encompassing the overall duration required for a program or task to finalize its execution. Latency plays a pivotal role in execution time, contributing to the total time invested in completing a given operation. For instance, in network communication, latency manifests as the time it takes for data to traverse from a source to a destination, influencing the overall execution time of tasks dependent on network interactions. Similarly, in computational processes, the latency associated with accessing data or resources contributes to the overall execution time, underscoring the interdependence of these two concepts.

Efforts to enhance system performance invariably involve addressing both latency and execution time. Optimizing algorithms, adopting efficient data structures, and mitigating communication delays are common strategies employed to minimize latency and, consequently, reduce overall execution time. This dual focus on latency and execution time optimization is essential for achieving responsive and efficient computing systems across various domains, from database queries to web applications, where minimizing delays is critical for delivering a seamless user experience and efficient task execution.

Some of the most notable metrics used to evaluate latency and execution time include \textbf {Average Execution Time} that offers a comprehensive perspective on service performance. Nevertheless, solely depending on the average might overlook the intricacies within the latency value distribution. \textbf{Median Execution Time} that is valuable for assessing the central point of the latency distribution, as it is not sensitive to outliers. A significant difference between the median and the average may indicate outliers disproportionately impacting latency. \textbf{Standard Deviation of Execution Time} that indicates the spread of execution times, with a higher standard deviation suggesting greater variability in latency. Monitoring standard deviation helps identify consistency or inconsistency in response times, crucial for user experience and detecting potential issues in the service infrastructure. \textbf{Maximum Execution Time} that represents the longest duration for task completion within a system, serving as an upper limit on acceptable execution time. Lower maximum execution time is desirable for timely and responsive performance, especially in real-time or time-sensitive applications. \textbf{Skewness of Latency} that assesses the asymmetry of latency distribution. A right-skewed distribution indicates some requests experience significantly longer delays than average, guiding optimizations to mitigate outliers. \textbf{Kurtosis of Latency} that measures the tails of the distribution, with higher kurtosis suggesting heavy tails and the presence of extreme values. Understanding kurtosis helps anticipate and manage rare but impactful events affecting service latency. \textbf{Tail Latency} (98th percentile) that focuses on extreme values in latency distribution, identifying the $2\%$ of requests with the longest response times. Monitoring tail latency ensures even under adverse conditions, a small percentage of users do not experience unacceptably long delays, directly impacting user satisfaction and SLAs.

In combination, these metrics are capable of serving as good indicators in the frame of service latency. For instance, high Average or Median Execution Times, coupled with high Maximum Execution Time, Skewness of Latency, and Kurtosis of Latency may indicate performance issues that need attention. Conversely, a low Average or Median Execution Time, combined with low Maximum Execution Time and well-behaved Skewness of Latency and Kurtosis of Latency suggests a more stable and predictable service. By regularly analyzing and interpreting these metrics, service providers can identify areas for improvement, and optimize performance, accordingly. However, as these metrics were formulated based on a relatively generic and broad definition of traditional monolithic services, they do not address the complexities inherent in modern services and distributed computing paradigms. Consequently, their capacity to identify areas for enhancement is constrained, and they fall short of providing a comprehensive evaluation of service performance within the context of contemporary services and distributed computing paradigms.

More specifically, the aforementioned metrics are unable to provide the two following insights regarding service performance:
\begin{itemize}
\item How often does latency surpass the specified threshold defined by the corresponding SLAs?
\item Given that the aforementioned threshold is surpassed, how long does it take for latency to become acceptable again? 
\end{itemize}

These two insights reflect the ability of a system to provide deterministic latency (beyond a specified threshold) as consistently as possible. Such an ability is especially important in the frame of contemporary services that aim to provide immersive experiences. The importance of having access to this type of information regarding evaluating the performance of a latency-sensitive service shall be explored in greater detail in Section \ref{exp}.

\section{Proposed Solution}
\label{prop}
In order to mitigate the aforementioned limitation, the authors of this work propose a new approach for evaluating the performance of contemporary latency-sensitive services. The proposed solution is based on the use of various fault tolerance metrics. Fault tolerance metrics are crucial for assessing the resilience and reliability of a system, particularly in the context of how well it can handle and recover from failures. Fault tolerance is a critical attribute in computing systems, reflecting the capacity to sustain normal functionality despite the occurrence of faults or errors. It involves the implementation of strategies and mechanisms to either prevent faults from causing system failures or to swiftly recover from failures when they do occur. The primary goal is to ensure uninterrupted service delivery and operational continuity, especially in mission-critical applications where system downtime or data loss could have severe consequences. By incorporating redundancy, error detection, and recovery mechanisms, fault-tolerant systems enhance overall reliability, providing users with a seamless experience and organizations with the assurance of continuous, dependable operations. The key fault tolerance metrics \cite{grecu2007essential} include Mean Time Between Failure (MTBF), Mean Time to Repair (MTTR), and Availability. Furthermore, MTBF, and MTTR can be leveraged to construct two additional metrics that are referred to as Reliability, and Maintainability.

\textbf{Mean Time Between Failures (MTBF)} is a critical metric in fault tolerance, representing the average duration between consecutive failures of a system. It serves as a key indicator of system reliability, with a higher MTBF value signifying a more dependable system. MTBF is calculated as the total Operational Time (the total time during which no failures occurred) divided by the Number of Failures during that time: $MTBF = \frac{Operational\ Time}{Number\ of\ Failures}$.

\textbf{Mean Time to Repair (MTTR)} is the average time it takes to restore a failed system or component to normal operation after a failure occurs. A lower MTTR is desirable, as it indicates faster recovery from failures, minimizing the impact on system availability and performance. MTTR is calculated as the Downtime (the total time during which failures emerged) divided by the Number of Failures: $MTTR = \frac{Downtime}{Number\ of\ Failures}$.

\textbf{Availability} is the percentage of time a system or service does not exhibit faults. The higher the value of Availability is, the  Availability is calculated as the total Operational Time divided by the sum of the Operational Time and the Downtime: $Availability = \frac{Operational\ Time}{Operational\ Time\ +\  Downtime}$.

\textbf{Reliability} is a measure of the probability that a system will operate without failure over a specified period. It provides an overall assessment of the system's dependability and consistency. A reliability value close to 1 indicates a highly dependable system, while values closer to 0 suggest a higher likelihood of failures. Reliability is calculated using the following formula: $Reliability = \frac{MTBF}{1\ +\ MTBF}$

\textbf{Maintainability} assesses how quickly and easily a system can be restored to operational status after a failure. Higher maintainability values indicate that the system is designed for efficient and effective maintenance, reducing the overall impact of failures on system performance. Maintainability is calculated using the following formula: $Maintainability = \frac{1}{1\ +\ MTTR}$

Much like the previously explored latency metrics, fault tolerance metrics exhibit various relations. Reliability and MTBF are closely related. A system with a high MTBF is likely to have high reliability, indicating that it can operate for an extended period without failure. Lower MTTR is often associated with higher maintainability. A system that can be quickly and easily repaired is more maintainable and, as a result, has a lower MTTR. The authors of this work advocate for the use of the aforementioned fault tolerance metrics in order to establish a more refined and detailed evaluation process in the frame of contemporary latency-sensitive services. More specifically, instead of using these metrics in the context of fault occurrence, they propose to use them on the basis of when latency exceeds a certain threshold. This approach is aligned with the use of specified thresholds, such as the ones that constitute the backbone of SLAs. 

According to the proposed approach, faults shall correspond to SLA violations that are defined by a specified latency threshold $t$. The Number of Failures shall correspond to the $Number\ of\ Violations(t)$ that emerge when task execution time exceeds the aforementioned threshold. Furthermore, the Operational Time shall correspond to $Time(No\ Violations(t))$ that is the time-span during which $t$ is not surpassed and Downtime shall correspond to $Time(Violations(t))$ that is the duration of time during which $t$ is surpassed. By using the $5$ fault tolerance metrics, one is capable of establishing $5$ new latency metrics that are based on the aforementioned formulation.

MTBF shall assist towards constructing \textbf{M1} that is calculated in the following manner: $M1 = \frac{Time(No\ Violations(t))}{Number\ of\ Violations(t)}$. MTTR shall assist towards constructing \textbf{M2} that is calculated in the following manner: $M2 = \frac{Time(Violations(t))}{Number\ of\ Violations(t)}$. Availability shall assist towards constructing \textbf{M3} that is calculated in the following manner: $M3 = \frac{Time(No\ Violations(t))}{Time(Violations(t))\ +\ Time(No\ Violations(t))}$. Reliability shall assist towards constructing \textbf{M4} that is calculated in the following manner: $M4 = \frac{M1}{1\ +\ M1}$. Maintainability shall assist towards constructing \textbf{M5} that is calculated in the following manner:  $M5 = \frac{1}{1\ +\ M2}$.

$M1$ and $M4$ reflect how frequently the latency surpasses the specified threshold defined by the corresponding SLA. $M1$ is calculated in seconds, while $M4$ is formulated as a fraction. $M2$ and $M5$ reflect how long it takes for latency to become acceptable again, after that the aforementioned threshold is surpassed. $M2$ is calculated in seconds, while $M5$ is formulated as a fraction. Finally, $M3$ is a fraction that indicates the percentage of time during which no SLA violations occur. Higher $M1$, $M3$, $M4$, $M5$, and lower $M2$ values indicate performance superiority in terms of latency. The formulation of the proposed metrics was established in accordance with the intricacies of distributed systems that consist of multiple computational nodes. Due to this fact, they acknowledge the fact that during a specific instance of time that is part of $Time(Violations(t))$, it is possible for multiple SLA violations to simultaneously emerge at different computational nodes of a distributed system. In other words, while $Time(Violations(t))$ and $Time(No\ Violations(t))$ are concepts that simultaneously encompass the entirety of the distributed system, the occurrence of SLA violations manifests at the level of computational nodes. Despite the fact that SLA violations occur at the level of computational nodes, they do not manifest in an independent manner since their emergence is indicative of the creation of bottlenecks in task execution that may affect other computational nodes as well. Finally, the transition from temporal instances of $Time(Violations(t))$ to temporal instances of $Time(No\ Violations(t))$ takes place only when there are no SLA violations across the entirety of the distributed system.

\section{Experimental Evaluation}
\label{exp}
Towards assessing the validity and usefulness of the proposed metrics in the frame of providing advanced insights into service performance we conducted a large-scale experiment on the basis of a horizontal autoscaling scenario using the CloudSim Plus \cite{cl} simulation framework. The simulation extended over a duration of $4$ days to accommodate periodic resource consumption phenomena spanning over multiple days. It involved the generation and offloading of over $500,000$ tasks to the available computational nodes to be processed. In the frame of the conducted experiment, we considered that there are $5$ computational nodes that are operating at all times, and that there are $15$ additional ones that can be invoked in order to handle increased resource demand. The tasks were generated using a combination of Poisson probability distributions and diverse statistical properties aligned with the task production rate present in contemporary latency-sensitive services, such as Multiplayer Mobile Gaming \cite{theodoropoulos2023graph}. To accurately capture the characteristics of service demand inherent in Multiplayer Mobile Gaming, application requests exhibited fluctuations throughout the day, escalating significantly after typical working hours at 17:00 (pm), and peaking at around 22:00 (pm). For the latency-related SLA threshold $t$ we have chosen $100$ milliseconds which is the de-facto upper limit for latency to be considered acceptable in gaming use-cases \cite{raaen2014latency}.

The purpose of the experiment was to comparatively analyse service performance when leveraging a reactive horizontal auto-scaling approach and a proactive one. The time required for acquiring additional resources was set to $5$ seconds, as a standard set by the Kubernetes Pod startup time \cite{kub}. In the case of the reactive scenario the allocation and de-allocation of computational resources is conducted on the basis of the ongoing CPU consumption. When CPU usage exceeds the $80\%$ threshold, the process of acquiring additional computational resources is triggered, while when CPU falls below $20\%$, the de-allocation process commences. In the case of the proactive scenario, the allocation and de-allocation of computational resources is conducted on the basis of the CPU consumption prediction that is made by a dedicated forecasting model. The prediction model receives the $6$ last recorded CPU usage values in order to generate a prediction that corresponds to the CPU usage that is expected to take place $10$ seconds into the future. Furthermore, it is based on a Deep Learning Encoder-Decoder \cite{theodoropoulos2021encoder} that is capable of performing accurate multi-step predictions. Finally, the forecasting model was created and executed using Python 3.9.13 and Tensorflow 2.9.1. Furthermore, the hardware configuration employed for both training and inference involved an i5-11400 CPU paired with an NVIDIA GeForce RTX 3060 GPU. 

The main idea behind the proactive approach is that by regulating resource scaling on the basis of the corresponding predictions, it is possible to provide notably improved Execution Time and Latency values by anticipating resource usage bottlenecks and allocating the optimal amount of resources in advance, thus circumventing the temporal restriction of the startup time that is associated with computational resources. One rather obvious means of mitigating Latency and Execution Time degradation is by increasing the number of computational resources that are being utilized \cite{theodoropoulos2023greenkube}. A proactive approach that leverages a forecasting model that overestimates future resource demand can provide reduced Latency and Execution Time values, regardless of the actual accuracy of the forecasting model. This may jeopardize the ability of the forecasting model to proactively detect sudden bursts in resource demand which can have dire ramification on service performance. Thus, the efficiency of the proactive approach highly depends on the accuracy of the forecasting model.

\subsection{Experimental Results \& Discussion}
Table \ref{tab1} depicts the experimental results that correspond to the proactive and the reactive approaches. Let us begin our analysis by considering only the conventional latency metrics. These include the Average Execution Time, the Median Execution Time, the Standard Deviation of Execution Time, the Maximum Execution Time, the Skewness of Latency, the Kurtosis of Latency, and Tail Latency. Based on the results that correspond to the Average, Median, and Maximum Execution Times, it is safe to conclude that the proactive approach manages to outperform the reactive one in terms of its overall ability to guarantee low latency in a generalised manner. Furthermore, the Standard Deviation of Execution time, Tail Latency, and Skewness \& Kurtosis of Latency values for the proactive approach indicate that while there is overall variability in latency values, extreme values or outliers are not as common across the majority of the distribution when compared against the reactive approach. Instead, these extreme values are concentrated in the tail, indicating the presence of occasional events that result in significantly higher latency. In other words, while the proactive approach is able to provide reduced latency when considering the entirety of the experiment, there are certain events during which latency is significantly increased compared to the reactive approach. This is the full extent of the insights that one can establish based on the use of these conventional latency metrics. Judging service performance solely based on the aforementioned metrics would indicate that the proactive approach is clearly the better solution.

\begin{table}[ht]
 \centering
    \begin{tabular}{l c c } \hline
    \toprule
        \multirow{2}{*}{\textbf{Metric}} & \multicolumn{2}{c}{\textbf{Approach}}  \\ \cline{2-3} 
         & \textbf{Reactive} & \textbf{Proactive} \\ \hline
\textbf{Average Ex. Time} & 1.482 & \textbf{1.435}  \\ 
\textbf{Median Ex. Time} & 1.459 & \textbf{1.249}   \\ 
\textbf{St. Dev. of Ex. Time} & \textbf{0.894} & 0.960   \\
\textbf{Maximum Ex. Time} & 14.790 & \textbf{14.229}  \\
\textbf{Skewness of Latency} & 3.967 & \textbf{3.671}   \\
\textbf{Kurtosis of Latency} & 26.120 & \textbf{20.654}   \\
\textbf{Tail Latency} & \textbf{4.194} & 4.516   \\
\hline \hline
\textbf{M1} & 1.754 & \textbf{2.217}  \\
\textbf{M2} & \textbf{0.878} & 0.980  \\
\textbf{M3} & 0.666 & \textbf{0.693} \\
\textbf{M4} & 0.636 & \textbf{0.689} \\
\textbf{M5} & \textbf{0.532} & 0.504 \\

\end{tabular}
\caption{Experimental results for the reactive and the proactive approaches using the conventional and the proposed latency metrics.} 
\label{tab1}
\end{table}

However, by examining the proposed latency metrics we are able to gain additional insights regarding service performance. In fact, by leveraging the proposed latency metrics we are able to find out that the proactive approach has some severe flaws that would otherwise remain unnoticed. Furthermore, we are able to identify the events that contribute towards the aforementioned increase in latency tails in the case of the proactive approach, and to even formulate an appropriate explanation for the reason why these events take place. The $M1$ and $M4$ values indicate that the proactive approach surpasses the proactive one in terms of the mean amount of time between SLA violations. As a result, during the proactive scenario SLA violations emerge less frequently. On top of that, the $M3$ values showcase that the proactive approach manages to present no SLA violations during a larger amount of time compared to the reactive approach. On the other hand, as the $M2$ and $M5$ values indicate, when they manifest, it takes a longer amount of time for latency values to return to acceptable levels, as defined by the SLA. Another important aspect that we have to consider is the fact that the $Number\ of\ Violations(t)$ and $Time(Violations(t))$ value were $21.47\%$ and $8.82\%$ higher for the reactive scenario. Consequently, if we consider a specific period of time across the two approaches, we see that during this specific period more SLA violations are likely to occur in the case of the reactive scenario. However, as stated before, in the case of the reactive approach, latency gets restored to acceptable levels in a more timely manner. This means that the higher $M2$ value of the proactive approach can not be attributed to the presence of a greater number of SLA violations that have a cascading effect on the time that is required for latency to get restored to acceptable levels. Instead, the worst $M2$ and $M4$ serve as strong indicators for the inability of the forecasting model to timely detect sudden future bursts in resource demand. This claim is also backed by the fact that the proactive approach utilizes $16.71\%$ more computational nodes compared to the reactive approach during the experiment. This indicates that the better Latency and Execution Time values are a direct result of the forecasting model's tendency to overestimate future resource usage which in turn triggers the allocation of a surplus of computational nodes. In the case of the reactive approach, SLA violations occur because the time required to startup a new computational node is equal to $5$ seconds, and as a result the system can not immediately invoke the allocation of additional computational nodes. On the other hand, in the case of the proactive approach, SLA violations occur due to the forecasting model's inability to timely detect sudden future bursts in resource demand. When an undetected surge of resource demand takes place, the forecasting model requires several sequential inputs that showcase increased CPU consumption before being capable of producing a prediction that encapsulates the aforementioned surge. Since new inputs are produced once every second, it may take several seconds to re-calibrate the forecasting model. Furthermore, even after the desired prediction is established, it takes $5$ additional seconds to allocate the requested computational nodes. The reactive approach, on the other hand, can acquire additional computational nodes within $5$ seconds.

\section{Conclusion}
\label{conclusion}

Conventional latency metrics, originally designed for traditional monolithic services, lack the adaptability required to comprehend the intricacies of modern services and distributed computing paradigms. As a result, their efficacy in pinpointing areas for enhancement is constrained, leaving a gap in delivering a thorough assessment of service performance within the dynamics of contemporary services and distributed computing paradigms. More precisely, these metrics fall short in shedding light on two crucial aspects of service performance: the frequency of latency surpassing predefined SLA thresholds and the time taken for latency to revert to an acceptable level post the threshold breach. This deficiency proves particularly significant in the realm of contemporary, latency-sensitive services, particularly immersive ones that demand consistent, deterministic low latency. In response to this limitation, the authors of this study introduce five innovative latency metrics. When utilized in conjunction with conventional latency metrics, these novel measures offer advanced insights capable of potentially enhancing service performance. The evaluation of the proposed metrics' validity and utility in furnishing advanced insights into service performance is conducted through an extensive large-scale experiment.

\section*{Acknowledgment}

This project has received funding from the EU’s Horizon 2020 program under Grant agreement No 101016509 (CHARITY). This paper reflects only the authors’ view and the Commission is not responsible for any use that may be made of the information it contains.

\bibliographystyle{IEEEtran}
\bibliography{references}

\end{document}